\documentclass[prl,twocolumn,showpacs,preprintnumbers,amsmath,amssymb,floatfix]{revtex4}

\usepackage{subfigure,graphicx,epsfig,amsmath,amsfonts,amssymb,xcolor,slashed,ulem,multirow}

\usepackage{relsize}
\usepackage{slashed}
\usepackage{color}
\usepackage{ulem}
\usepackage{tabu}

\usepackage{amsmath}
\usepackage{amssymb}
\usepackage{amsthm}
\usepackage{mathrsfs}
\usepackage{graphicx}
\usepackage{epstopdf}
\usepackage{fancyhdr}
\usepackage{array}
\usepackage[all]{xy}
\usepackage{eufrak}
\usepackage{euscript}
\usepackage{enumerate}
\usepackage{slashed}
\usepackage{hyperref}
\usepackage{subfigure} 
\usepackage{epstopdf} 

\usepackage{hyperref}
\hypersetup{pdftex,colorlinks=true,linkcolor=blue,citecolor=blue,menucolor=black,urlcolor=blue,filecolor=blue}

\newcommand{\qvp}{\vec{q}\,'}

\newcommand{\qmax}{q_{\rm max}}

\newcommand{\mev}{\textrm{ MeV}}

\newcommand{\be}{\begin{equation}}
\newcommand{\ee}{\end{equation}}
\newcommand{\ba}{\begin{eqnarray}}
\newcommand{\ea}{\end{eqnarray}}
\newcommand{\nn}{\nonumber}

\newcommand{\barb}{{\bar B}}
\newcommand{\bark}{{\bar K}}
\newcommand{\barbs}{{\bar B^*}}
\newcommand{\barks}{{\bar K^*}}
\newcommand{\bs}{B^*}

\begin{document}
\title{Predictions of super-exotic heavy mesons from $K^{*}B^{*}B^{*}$ interactions}

\author{M. Bayar}
\email{melahat.bayar@kocaeli.edu.tr}
\affiliation{Department of Physics, Kocaeli University, 41380, Izmit, Turkey}

\author{N. Ikeno}
\email{ikeno@tottori-u.ac.jp}
\affiliation{Department of Agricultural, Life and Environmental Sciences, Tottori University, Tottori 680-8551, Japan}
\affiliation{Cyclotron Institute, Texas A\&M University, College Station, Texas 77843, USA}

\author{L.~Roca}
\email{luisroca@um.es}
\affiliation{Departamento de F\'isica, Universidad de Murcia, E-30100 Murcia, Spain}

\preprint{}

\date{\today}

\begin{abstract}
 We make a theoretical study of the three-body system composed of $\barks\barbs\barbs$  to look for possible bound states, which could be associated to mesonic resonances of very exotic nature, containing open  strange and double-bottom flavours. The three-body interaction is evaluated by using the fixed center approach to the Faddeev equations where the $\barbs\barbs$ is bound forming an $I(J^P)=0(1^+)$ state, as it was found in previous works, and the third particle, the $\barks$, of much smaller mass, interacts with the components of the cluster. We obtain bound states for all the channels considered: spin $J=0$, 1 and 2, all of them with isospin $I=1/2$ and negative parity.
\end{abstract}


\maketitle

\section{Introduction}

The meson spectrum in the heavy flavor sector has gained a renewed impetus in the last two decades thanks to a significant increase of experimental results (see Ref.~\cite{Chen:2016spr,Brambilla:2019esw} for reviews). Of special interest and repercussion  has  been the proliferation of exotic states, which cannot be explained as ordinary  $q\bar q$ mesons, like the hidden heavy flavour  $XYZ$ resonances, with  theoretical interpretations  ranging from tetraquarks to molecular states \cite{olsenxyz,xliuxyz,Hosaka:2016pey,Chen:2016qju,Brambilla:2019esw}.
Even more challenging has been the recent discovery of non-$q \bar q$ open flavor mesons like the $X_0(2900)$ \cite{LHCb:2020bls,LHCb:2020pxc}
with an open charm and strange flavour, which is undoubtedly exotic since it contains at least a $c$ and an $s$ quark and then needs at least two other antiquarks to form a color singlet. 
The theoretical interpretations of the $X_0(2900)$ range from the picture of tetraquarks 
\cite{Wang:2020xyc,He:2020jna,Zhang:2020oze,Wang:2020prk} to a molecular structure \cite{Liu:2020nil,Chen:2020aos,Huang:2020ptc,Molina:2020hde,Dai:2021vgf,Xue:2020vtq,Agaev:2020nrc,Mutuk:2020igv,Xiao:2020ltm,He:2020btl} or even a kinematic  triangle singularity 
\cite{Liu:2020orv,Burns:2020epm}.
Specially sound has been the recent discovery of the manifestly exotic open double-charm $T_{cc}(3875)$
\cite{f4,f23} also with a natural interpretation as a molecular $D^*D$ state \cite{feijoo,f30,f32,du,miguel}.
The success of the molecular interpretation has triggered the search for possible bound states for other exotic heavy flavor combinations like open double-bottom from 
$\bs B$, $\bs\bs$, $\bs_s \bs$ interaction \cite{Dai:2022ulk}; open $c$ and $b$ from $\barb D$, $\barbs D$, $\barb D^*$, $\barb \bar D^*$ \cite{Sakai:2017avl}; and open $b$ and $s$ flavours from $B K$, 
$B^* K$, $B K^*$, $B^* K^*$ interaction \cite{Oset:2022xji}.
Most of these molecular interpretations are based on the implementation of unitarity to the amplitudes obtained from extensions of the lowest order chiral Lagrangians from where, in many cases, bound states and resonances appear dynamically without  the need to include them as explicit degrees of freedom (see \cite{Oller:2000ma} for a classical early review and \cite{Guo:2017jvc,Dong:2021bvy,Dong:2021juy} for recent reviews of results in the heavy sector).

A natural and challenging step forward is to consider the extension to three  body systems.  In the last decade dozens of works have found many states theoretically, even with open or hidden heavy flavours (see Ref.~\cite{MartinezTorres:2020hus} for a review and list of references).
The three body system allows for even  the possibility to have super exotic mesons with three open flavors  like for instance $bbb$, which was found to bind in the $B \bs\bs-\bs\bs\bs$ interaction in \cite{Garcilazo:2018rwu}; or $ccs$, where bound states were found for several total spins in the
$D^*D^*\barks$ system in \cite{Ikeno:2022jbb}.
These states would require at least three extra antiquark to get the color singlet, then would correspond to hexaquarks in the standard quark picture.
In the present work we take advantage of the findings of this latter work and extend the formalism to the $\barks\barb^*\barb^*$ interaction to look for possible bound states which would have  open $bbs$ flavours.
The $\barks\barb^*\barb^*$ system is very interesting since it is significantly different from the $\bar{K}B^* B^*$~\cite{Valderrama:2018knt} and $\bar{K}^{(*)}B^{(*)}\bar{B}^{(*)}$~\cite{Ren:2018qhr} systems with  hidden bottom flavours.

The standard way to tackle the three-body scattering problem has traditionally been to try to solve the Faddeev 
equations  \cite{Faddeev:1960su} implementing approximate methods, due to the practical impossibility to solve them exactly.
This is indeed a well known problem in  nuclear and hadron physics 
like in the three-nucleon 
interaction \cite{Alt,Epelbaum}, systems involving baryons and mesons 
\cite{nogami,Ikeda:2007nz,MartinezTorres:2007sr,Jido:2008kp} or three-meson 
interaction \cite{Mennessier:1972bi,MartinezTorres:2008gy,MartinezTorres:2009xb}.
However, when two of the three particles are strongly correlated among themselves, and the third particle is lighter than the other particles  \cite{MartinezTorres:2010ax}, the Faddeev equations
can be strongly simplified and one can make use of a formalism called Fixed 
Center Approximation  (FCA) to the Faddeev equations 
\cite{Chand:1962ec,Barrett:1999cw,Deloff:1999gc,Kamalov:2000iy,Gal:2006cw}.
The FCA have been successfully applied to dozens of three-body systems (see  Table~1 in Ref.~\cite{MartinezTorres:2020hus} for a list of different works).
 It is worth mentioning here that in related problems the FCA has been compared to the variational method and similar results have been found. This is the case of the $D\bar D K$ system studied in \cite{Wu:2020job} with the variational method and in 
\cite{Wei:2022jgc} with the FCA, or the case of the $D^* D^*D^*$ system studied in \cite{Luo:2021ggs} with the variational method and in \cite{Bayar:2022bnc} with the FCA.

In the present work we will study the $\barks\barbs\barbs$ within the FCA, because in a previous work \cite{Dai:2022ulk} it was found that the $\barbs\barbs$ in $I(J^P)=0(1^+)$ was bound with a binding energy of about 40~MeV. In addition, the $\barks\barbs$ was also found to be strongly attractive in \cite{Oset:2022xji} for all possible spins in $I=0$. Then we can expect with confidence that the three body $\barks\barbs\barbs$ will present bound states. It adds even more confidence the fact that bound states were found in \cite{Ikeno:2022jbb} in the $D^*D^*\barks$ system, where analogously to our case the $D^*D^*$ is bound \cite{Dai:2021vgf} and the $D^* \bar{K}^*$ is also attractive.
Advancing some results, we find three-body states for all the three spin channels, $J=0$, 1 and 2.

\section{Formalism}

\subsection{Three-body scattering}

The FCA to the Faddeev 
equations is an effective way to evaluate the three-body scattering when two of the particles form a bound state, which will be called cluster, 
 and it is not excited in the intermediate states \cite{MartinezTorres:2010ax}. If the third particle is much lighter than the constituents of the cluster  it is unlikely to have enough  available energy to excite it.
 In the present case, $\barks\barb^*\barb^*$, we are in this situation. Indeed in Ref.~\cite{Dai:2022ulk} it was obtained, among other states,  that the $\barbs\barbs$ system in $I(J^P)=0(1^+)$ was bound with about  40~MeV, and the third particle of the three-body system is a $\barks$, which is much lighter than the $\barbs$ making up the cluster.
  The projectile, $\barks$, rescatters repeatedly with each component of the cluster.
 \begin{figure}[!t]
\begin{center}
\includegraphics[width=0.99\linewidth]{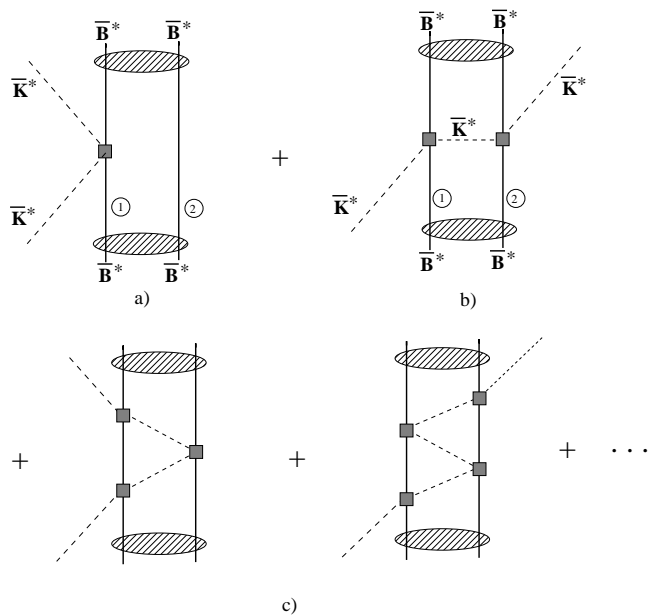}
\caption{Representation of the fixed center approximation
to the Faddeev equations for the interaction of a $\barks$ meson with a $\barbs\barbs$ bound state.
 Diagrams $a)$ represents the single scattering contribution,  $b)$ the double scattering one and $b)+c)$ the
multiple scattering.}
\label{fig:FCA}
\end{center}
\end{figure} 
   This is depicted diagrammatically in Fig.~\ref{fig:FCA},
   and the total three-body scattering amplitude, $T$, can then be formally written as a system of coupled equations
  
  \begin{align}
  T_1&=t_1+t_1G_0T_2\nn\\
  T_2&=t_2+t_2G_0T_1\nn\\
  T&=T_1+T_2
  \label{eq:FCA1}
   \end{align}
\noindent
where the two partition functions, $T_i$, account for all the diagrams starting with the interaction of the $\barks$ with the i-th $\barbs$ particle in the cluster. In the present case, since the two particles in the clusters are the same, we have $t_1=t_2$, $T_1=T_2$, and, hence Eq.~\eqref{eq:FCA1} decouples.
Diagram Fig.~\ref{fig:FCA}a) represents the single-scattering, $t_1$, and Fig.~\ref{fig:FCA}b) the double scattering contribution, $t_1G_0t_2$. The infinite sum in Fig.~\ref{fig:FCA}c) represents the rest of the rescattering to get the full $T_1$ amplitude in the FCA.
In Eq.~\eqref{eq:FCA1}, $G_0$, stands for the Green function representing the exchange of a $\barks$ between the $\barbs$ mesons inside the compound system, and is represented by a dashed line in Fig.~\ref{fig:FCA}, and which is given by \cite{Roca:2010tf,YamagataSekihara:2010qk, Roca:2011br}
\begin{align}\label{G0}
G_0(q^0)=\frac{1}{2M_c}\int\frac{d^3\vec{q}}{(2\pi)^3}\frac{F(\vec{q}\,)}
 {(q^{0})^2-\omega_{K^*}^2(\vec{q}\,)+i\epsilon} \,\, ,
\end{align}
with $\omega_{K^*}(\vec{q}\,)=\sqrt{|\vec{q}\,|^2+m_{K^*}^2}$. In Eq.~\eqref{G0}, $q^0$ is the  energy carried 
by the $\barks$ meson between the components of the cluster, given by
\begin{align}
 q^0=\frac{1}{2M_c}(s-m_{K^*}^2-M_c^2)
\end{align}
and $M_c$ is the mass of the  $\barbs\barbs$ bound state (cluster) from Ref.~\cite{Dai:2022ulk}, the value of which is explained in the results section.

The form factor $F(\vec{q}\,)$ in Eq.~\eqref{G0} encodes the information  about the $\barbs\barbs$ bound state, which is related to the 
cluster wave function, $\Psi_c(\vec r\,)$, by means of a Fourier transform \cite{Roca:2010tf,YamagataSekihara:2010pj}, 
\begin{align}
F(\vec q\,)=\int d^3\vec r \, e^{-i\vec q\cdot\vec r} \Psi_c^2(\vec r\,).
\end{align}
The form factor can be derived in a similar way as done in \cite{Roca:2010tf,YamagataSekihara:2010pj} and gives
\begin{align}\label{eq_ff}
 F(\vec{q}\,)=\frac{1}{N}\int_\Omega
 d^3\qvp\frac{1}{M_c-2\omega_{\barbs}(\qvp)}
 \frac{1}
 {M_c-2\omega_{\barbs}(\vec{q}-\qvp)}\, ,
\end{align}
where $\Omega$ specifies the conditions $|\qvp|<\qmax$ and $|\vec{q}-\qvp|<\qmax$.
 The normalization factor $N$ in Eq.~\eqref{eq_ff} guarantees that $F(\vec{q}=0)=1$, and thus it is given by

\begin{eqnarray}\label{norm}
N = \int\limits_{|\vec{q'}\,|<\qmax}\, d^3 \vec{q'}\, \Big( \frac{1}
{M_c  - 2\omega_{\barbs}(\vec{q'}\,) } \Big)^2 \, .
\end{eqnarray}

\begin{figure}[!t]
\begin{center}
\includegraphics[width=1\linewidth]{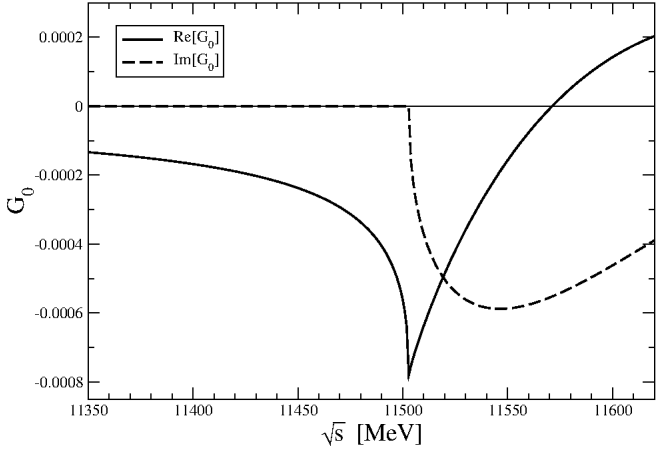}
\caption{Real and imaginary parts of the $G_0$ function, 
Eq.~(\ref{G0})}
\label{fig:G0}
\end{center}
\end{figure}

In the results section we discuss the value used for the cutoff $\qmax$ in the three-momentum integration.
 \begin{figure}[!t]
\begin{center}
\includegraphics[width=0.8\linewidth]{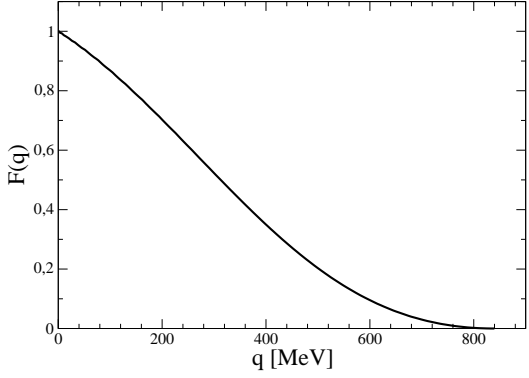}
\caption{Form factor of the $\barbs\barbs$ bound state}
\label{fig:Ff2}
\end{center}
\end{figure}
In Fig.~\ref{fig:Ff2} we show the form factor as a function of the modulus of the momentum for $\qmax=420\mev$ and in Fig.~\ref{fig:G0}
 the real and imaginary parts of the $G_0$ function, which close to threshold resembles very much the typical shape of the two meson loop function,  in this case the $\barks$ and the meson made of two $\barbs$.

Finally, there is an important issue regarding the normalization of the amplitudes that one has to take into account when mixing, in the same expression, Eq.~\eqref{eq:FCA1}, three-body amplitudes, $T$,  with two-body ones.
Using the Mandl-Shaw \cite{MandlShaw:2010} normalization for the mesonic fields, the $S$ matrix for the single scattering
contribution can be written as:
\ba
S^{(1)}=S^{(1)}_1+S^{(1)}_2  ,
\ea
with
\ba
&&S^{(1)}_i=-it_{Ab_i} \frac{1}{{\cal V}^2}
\frac{1}{\sqrt{2\omega_{p_i}}}
\frac{1}{\sqrt{2\omega_{p'_i}}}
\frac{1}{\sqrt{2\omega_k}}
\frac{1}{\sqrt{2\omega_{k'}}}\nonumber\\
&&\times(2\pi)^4\,\delta(k+k_{c}-k'-k'_{c})
.
\label{eq:single}
\ea
where 
$t_{Ab_i}$ are the single scattering two-body amplitudes,
$\cal V$ is an irrelevant normalization volume, 
$\omega_p=\sqrt{p^2+m^2}$ is the on-shell energy of a
particle with momentum $p$ and mass $m$,
$p_i$ ($p'_i$) is the initial (final) momentum of the particle $b_i$ in the cluster, $k$ ($k'$) represent the initial (final) momentum of the projectile $A$ and $k_c$ ($k'_c$)
represents the total momentum of the
initial (final) cluster. 
On the other hand, the general form
of the $S$-matrix of the three-body interaction is 
\ba
S&=&-i T(2\pi)^4 \delta(k+k_c-k'-k'_c)\frac{1}{{\cal V}^2}\nn\\
&&\times\frac{1}{\sqrt{2 \omega_k}} 
\frac{1}{\sqrt{2 \omega_{k'}}}
\frac{1}{\sqrt{2 \omega_{k_c}}}
\frac{1}{\sqrt{2 \omega_{k'_c}}},
\label{eq:finalS2AB}
\ea
and comparing this equation with Eq.~(\ref{eq:single}), we 
get that  the 
FCA equations (\ref{eq:FCA1})  take  the form
\begin{eqnarray}
\bar T_1&=& \bar t_1 + \bar t_1 G_0 \bar T_2  \nonumber\\
\bar T_2&=& \bar t_2  + \bar t_2 G_0 \bar T_1\nonumber\\
\bar T&=&\bar T_1+\bar T_2
\label{eq:Faddeevsystem}
\end{eqnarray}
with
\be \bar t_i=  \sqrt{\frac{\omega_{k_c} \omega_{k_c'}}{\omega_{p_i}
 \omega_{p_i'}}}
t_{A b_i}(s_i),
\ee
which in our case can be approximated by
\ba
\bar t_1=\bar t_2=\frac{M_c}{m_\barbs}t_1.
\ea

With all these ingredients, Eq.~\eqref{eq:Faddeevsystem} can be algebraically solved and gives, for the total three body-amplitude,

\ba
\bar T= 2 \bar T_1= \frac{2}{\bar t_1^{-1}-G_0}
\label{eq:barTfinal}
\ea

\subsection{Two-body interaction}

For the evaluation of the two-body $t_i$ amplitudes, represented by the full squares in Fig.~\ref{fig:FCA}, we need the $\barks\barbs$ interaction, which amplitudes are obtained from Ref.~\cite{Dai:2022ulk}. Note that the amplitudes in Ref.~\cite{Oset:2022xji} are provided for a given isospin and spin, therefore we have to write the total isospin state of the global system in terms of the coupled isospin state of the $\barks$ and a $\barbs$ of the cluster.
For a general situation where the incident particle is called $A$ and the cluster $B$ is made of two particles, $b_1$ and $b_2$,
the amplitudes $t_i$ in Eq.~\eqref{eq:FCA1} actually stand for matrix  elements between the eigenstates

\ba
& &| I_A,I_B,I,M  \rangle
\ea
where $I_A$ is the isospin of the particle $A$, $I_B$ the isospin of the cluster
$B$, $I$ the total $AB$ isospin and $M$ the third component of the total isospin $I$.
This state must then be first written in terms of the ket
\ba
& &| I_A,I_i,I_{Ai},M_{Ai}  \rangle
\ea
where $I_i$ is the isospin of particle $b_i$ and 
 $I_{Ai}$ the global isospin of the $A$-$b_i$ system. In order to do this, one can first write $| I_A,I_B,I,M  \rangle$ in terms of 
$| I_A,M_A;I_B,M_B  \rangle$, then write $| I_B,M_B  \rangle$ in terms of
$| I_1,M_1,I_2,M_2  \rangle$ and finally 
$| I_A,M_A;I_i,M_i  \rangle$ in terms of $| I_A,I_i,I_{Ai},M_{Ai}  \rangle$.
Thus the  expression of $| I_A,I_B,I,M  \rangle$ in terms of $|I_A,I_i,I_{Ai},M_{Ai}\rangle\otimes|I_j, M-M_{Ai}\rangle$ is \cite{Roca:2011br}
\ba
& &| I_A,I_B,I,M  \rangle^{(i)} =\nn \\
& &\sum_{I_{Ai}}\sum_{M_{Ai}}\sum_{M_A}
{\cal C}(I_A,I_B,I|M_A,M-M_A,M)\nn \\ 
&&\times
{\cal C}(I_i,I_j,I_B|M_{Ai}-M_A,M-M_{Ai},M-M_A)\nn \\ 
&&\times
{\cal C}(I_A,I_i,I_{Ai}|M_A,M_{Ai}-M_A,M_{Ai})\nn \\ 
&&\times|I_A,I_i,I_{Ai},M_{Ai}\rangle\otimes|I_j, M-M_{Ai}\rangle
\label{eq:stateclebsh}
\ea 
where 
the superscript $(i)$ indicates that we are correlating the particle $A$ with $b_i$ and
${\cal C}(j_1,j_2,j_3|m_1,m_2,m_3)$  represent Clebsch-Gordan coefficients.

In the present case one has to consider that the $\barbs\barbs$ cluster is in isospin 0 ($I_B=0$), that the total three-body isospin is $I=1/2$, and that the isospin doublets are $(\bar{K}^{*0},-\bar{K}^{*-})$ and $(\bar{B}^{*0},-\bar{B}^{*-})$. Then using, Eq.~\eqref{eq:stateclebsh}, we have
\ba
&|\barks(\barbs\barbs)\rangle^{(1)}_{I=1/2,M=1/2} 
=  -\frac{1}{2}|\barks\barbs\rangle_{I=0,M=0}|\bar{B}^{*0}\rangle \nn\\
&-\frac{1}{2}|\barks\barbs\rangle_{I=1,M=0}|\bar{B}^{*0}\rangle
-\frac{1}{\sqrt{2}}|\barks\barbs\rangle_{I=1,M=1}|\bar{B}^{*-}\rangle. \nn\\
\label{eq:stateKBB}
\ea

The amplitude for the single scattering contribution can be
written in terms of the two body amplitudes,  $t^{(I_{Ai})}_{Ab_i}$,
for the transition $Ab_i\to Ab_i$ with isospin $I_{Ai}$:

\ba
&&\langle  I_A,I_B,I,M | t_i |  I_A,I_B,I,M \rangle
=\sum_{I_{Ai}}\bigg[\sum_{M_{Ai}}\sum_{M_A}\sum_{M_A'}\nn \\ 
&&\times
        {\cal C}(I_A,I_B,I|M_A,M-M_A,M)\nn \\ 
&&\times{\cal C}(I_A,I_B,I|M_A',M-M_A',M)\nn \\ 
&&\times
        {\cal C}(I_i,I_j,I_B|M_{Ai}-M_A,M-M_{Ai},M-M_A)\nn \\ 
&&\times
        {\cal C}(I_{i},I_{j},I_{B}|M_{Ai}-M_A,M-M_{Ai},M-M_{A'})\nn \\ 
&& \times
        {\cal C}(I_A,I_i,I_{Ai}|M_A,M_{Ai}-M_A,M_{Ai})\nn \\ 
&&\times{\cal C}(I_{A},I_{i},I_{Ai}|M_{A'},M_{Ai}-M_{A'},M_{Ai})\bigg]
 \times
\ t^{(I_{Ai})}_{Ab_i}\nn \\ 
& \equiv& \sum_{I_{Ai}} \alpha_{i}\ t^{(I_{Ai})}_{Ab_i}.
\label{eq:V121p2p}
\ea
which is easily implementable for computer evaluation for a general case and thus it is why we quote it here, since it may be useful in other works. In our case it is more direct to obtain it from Eq.~\eqref{eq:stateKBB},
\ba
t_1= \frac{1}{4} t_{\barks\barbs}^{I=0} + 
\frac{3}{4} t_{\barks\barbs}^{I=1}.
\label{eq:t1I}
\ea 

 On the other hand, we also have to consider the different spin combinations and write the total three-body spin amplitudes in terms of the spin of the two-body $\barbs\barbs$. The reasoning is then totally analogous to the previous discussion about the isospin and hence we can use the master formula Eq.~\eqref{eq:V121p2p} but changing isospin by spin. Then, using the actual spins of the particles involved and taking into account that the bound $\barbs\barbs$ state has $J=1$, we get, for the different possible values of the total spin, $J=0,1,2$,
 
 \ba
 t^{J=0}_1&=& t_{\barks\barbs}^{J=1} \nn \\
 t^{J=1}_1&=& \frac{1}{3}t_{\barks\barbs}^{J=0}
  +\frac{1}{4}t_{\barks\barbs}^{J=1} 
  + \frac{5}{12}t_{\barks\barbs}^{J=2} \nn \\
 t^{J=2}_1&=&\frac{1}{4}t_{\barks\barbs}^{J=1} 
  + \frac{3}{4}t_{\barks\barbs}^{J=2}.
  \label{eq:t1J} 
 \ea
 Combining equations \eqref{eq:t1I} and \eqref{eq:t1J} we finally get
  \begin{align}
 t^{J=0}_1&= 
  \frac{1}{4} t_{\barks\barbs}^{(I=0,J=1)}
  + \frac{3}{4} t_{\barks\barbs}^{(I=1,J=1)}
  \nn \\
  t^{J=1}_1&= 
  \frac{1}{12} t_{\barks\barbs}^{(I=0,J=0)}  
+ \frac{1}{16} t_{\barks\barbs}^{(I=0,J=1)} 
+ \frac{5}{48} t_{\barks\barbs}^{(I=0,J=2)}\nn \\
&+ \frac{1}{4} t_{\barks\barbs}^{(I=1,J=0)} 
+ \frac{3}{16} t_{\barks\barbs}^{(I=1,J=1)}
+ \frac{5}{16} t_{\barks\barbs}^{(I=1,J=2)} \nn \\
 t^{J=2}_1&= 
 \frac{1}{16} t_{\barks\barbs}^{(I=0,J=1)}
+ \frac{3}{16} t_{\barks\barbs}^{(I=0,J=2)}\nn \\
&+ \frac{3}{16} t_{\barks\barbs}^{(I=1,J=1)}
+ \frac{9}{16} t_{\barks\barbs}^{(I=1,J=2)} 
  \label{eq:t1IJ}   
 \end{align}
Note that the
 argument of the function $T_i$ in Eq.~\eqref{eq:FCA1} is
  the total invariant mass energy, $s$, of the three-body system. However the argument  of the two-body $\barks\barbs$ amplitudes in Eq.~\eqref{eq:t1IJ}, and hence $t_1$ and $t_2$ in Eq.~\eqref{eq:FCA1},  are $s_1$ and $s_2$, where
$s_i (i=1,~2)$ is the invariant mass of the interacting particle $A$ and
the particle $b_i$ of the $B$ molecule and is given by \cite{YamagataSekihara:2010qk}

\begin{equation}
s_i=m_A^2+m_{b_i}^2+\frac{1}{2 m_B^2}(s-m_A^2-m_B^2)
(m_B^2+m_{b_i}^2-m_{b_{j\ne i}}^2),
\label{eq:si}
\end{equation}
which in our case gives
\begin{equation}
s_1=s_2=m_{K^*}^2+m_{B^*}^2+\frac{1}{2}(s-m_{K^*}^2-M_c^2)
,
\label{eq:s1}
\end{equation}
where $M_c$ is the mass of the  $\barbs\barbs$ bound state.

The $t_{\barks\barbs}$ amplitudes for $I=0$
 are obtained from Ref.~\cite{Oset:2022xji}
by implementing unitarity by means of the Bethe-Salpeter equation, starting with 
potential kernels, $V$, obtained from the dominant vector meson exchange interaction plus four vector contact interaction:
\be
t_{\barks\barbs}=[1-VG_{\barks\barbs}]^{-1}V\,.
\label{eq:TBS}
\ee
 The elementary
vertices in the evaluation of $V$ are supplied by local hidden gauge
symmetry Lagrangians properly extended to the bottom sector.
In this model, the $\barks\barbs$ scattering amplitudes present poles for  $I(J^P)=0(0^+)$, 
$0(1^+)$ and $0(2^+)$ with binding energies of the order of 100~\mev.
In Ref.~\cite{Oset:2022xji}
the widths of the generated states were also evaluated by identifying  the main sources of imaginary part, which turned out to be the width of the $\barks$  and the box diagrams with intermediate  $\bark\barb$ and $\bark\barbs$ states. (See details of the formalism and calculations in Ref.~\cite{Oset:2022xji}). 

The $t_{\barks\barbs}$ amplitudes in $I=1$ are not calculated in 
Ref.~\cite{Oset:2022xji} and thus we evaluate them in the present work considering the same contact term and exchange of $\rho$, $\omega$ and $B_s^*$ mesons. The contact term contribution is
\be
V^{I=1}_\textrm{contact}= \left\{\begin{array}{cc}
-4 g^2&\qquad\textrm{for $J=0$},\\
0  &\qquad\textrm{for $J=1$},\\
2 g^2&\qquad\textrm{for $J=2$}.
\end{array}\right.
\label{eq:tcontact}
\ee
where $g=800\mev/(2f_\pi)$, with $f_\pi=93\mev$,
and
\begin{align}
V^{I=1, J=0,2}_\textrm{exch.}&= 
\frac{g^2}{m^2_{B_s^*}}(p_1+p_4)(p_2+p_3)\nn\\
+&\frac{1}{2}g^2\left(\frac{1}{m_\omega^2}+\frac{1}{m_\rho^2}
\right)(p_1+p_3)(p_2+p_4),\\
V^{I=1, J=1}_\textrm{exch.}&= -\frac{g^2}{m^2_{B_s^*}}(p_1+p_4)(p_2+p_3)\nn\\
+&\frac{1}{2}g^2\left(\frac{1}{m_\omega^2}+\frac{1}{m_\rho^2}
\right)(p_1+p_3)(p_2+p_4),
\label{eq:VBsKsJ}
\end{align}
where we have to carry out an s-wave projection of the momentum structures which gives \cite{Roca:2005nm,Oset:2022xji}
\begin{align}\label{eq:p1p3p2p4}
(p_1+p_3)(p_2+p_4)&\to \frac{1}{2}\big[3s-(m_1^2+m_2^2+{m_3}^2+{m_4}^2).\nn\\
&-\frac{1}{s}(m_1^2-m_2^2)
({m_3}^2-{m_4}^2)\big]\,,\\
(p_1+p_4)(p_2+p_3)&\to \frac{1}{2}\big[3s-(m_1^2+m_2^2+{m_3}^2+{m_4}^2)\nn\\
&+\frac{1}{s}(m_1^2-m_2^2)
({m_3}^2-{m_4}^2)\big]\,,
\label{eq:p1p4p2p3}
\end{align}
with $m_i$ the mass of the particle with momentum $p_i$, of the process $\barks(p_1)\barbs(p_2)\to\barks(p_3)\barbs(p_4)$.
This potential at threshold takes the value $29g^2$ for $J=0$, 
$30 g^2$ for $J=1$ and $35g^2$ for $J=2$, which are strongly repulsive and then the $I=1$ 
$\barks\barbs$ amplitudes do not develop bound states unlike the $I=0$ ones when implementing the unitarization procedure through the Bethe-Salpeter equation.

\section{Results}

For the numerical evaluation, the three-momentum cutoff $\qmax$ of Eq.~\eqref{eq_ff} is, in principle, a free parameter of the model. However 
it is conceptually analogous to the regularization cutoff used in the 
calculation of the $\barbs\barbs$ loop function needed in the Bethe-Salpeter equation to obtain the  $\barbs\barbs$ bound state in
Ref.~\cite{Dai:2022ulk}. 
Indeed it was shown in Ref.~\cite{Gamermann:2009uq} that 
the use of a separable two-body potential in momentum space with a maximum momentum $\qmax$, $V=v\theta(\qmax -q)\theta(\qmax -q')$,
with $q$ $(q')$ the modulus of the initial (final) scattering momenta,  converts the
 coupled  integral Bethe-Salpeter equation into an algebraic one with on-shell prescriptions and the $\qmax$ translate into the cutoff of the loop function as in Eq.~\eqref{eq_ff}.
 Therefore we use the same value $\qmax\in[400,450]$ as was used for the evaluation of the $\barbs\barbs$ loop in Ref.~\cite{Dai:2022ulk}.
The values obtained in  Ref.~\cite{Dai:2022ulk} for the mass of the 
 $\barbs\barbs$ bound state were $10612\mev$ and $10607\mev$ for 
 $\qmax=400\mev$ and $\qmax=450\mev$ respectively.
 On the other hand, in the model for the two-body $\barks\barbs$ amplitudes \cite{Oset:2022xji} there was also an uncertainty from the value of the cutoff used in the $\barks\barbs$ loop function. This cutoff was obtained in that work by fitting the  experimental mass of the $X_0(2866)$ state, obtained in the $\barks D^*$  interaction. 
The use of the same value for the cutoff for $\barks\barbs$ than for $\barks D^*$ is justified within the heavy quark spin symmetry, since the value of the cutoff is
independent of the heavy quark flavor, up to corrections of order ${\cal O}(1/m_Q)$ with $m_Q$ the mass of the heavy quark \cite{Nieves:2011zz}. This value of the $\barks\barbs$ cutoff found in \cite{Oset:2022xji} was $1050\mev$, which we will call $\qmax^{\barks\barbs}$ in the following to distinguish it from the 
 $\qmax$ of the $\barbs\barbs$ regularization and form factor described above. For the estimation of uncertainties, a range between $900\mev$ and $1050\mev$ was used in  \cite{Oset:2022xji} and then we will also use that range for $\qmax^{\barks\barbs}$ in the present work. The reason for the consideration also of the lower value of $\qmax^{\barks\barbs}$ is that one expects the value of the cutoff to be of the order of the inverse of the range of the interaction, and since the potential is dominated by vector meson exchange one expects to be closer to the mass of the vector mesons considered.
 The value of $\qmax^{\barks\barbs}$ is the largest source of uncertainty of the present calculation.

\begin{figure}[!ht]
     \centering
      \subfigure[]{
          \includegraphics[width=.88\linewidth]{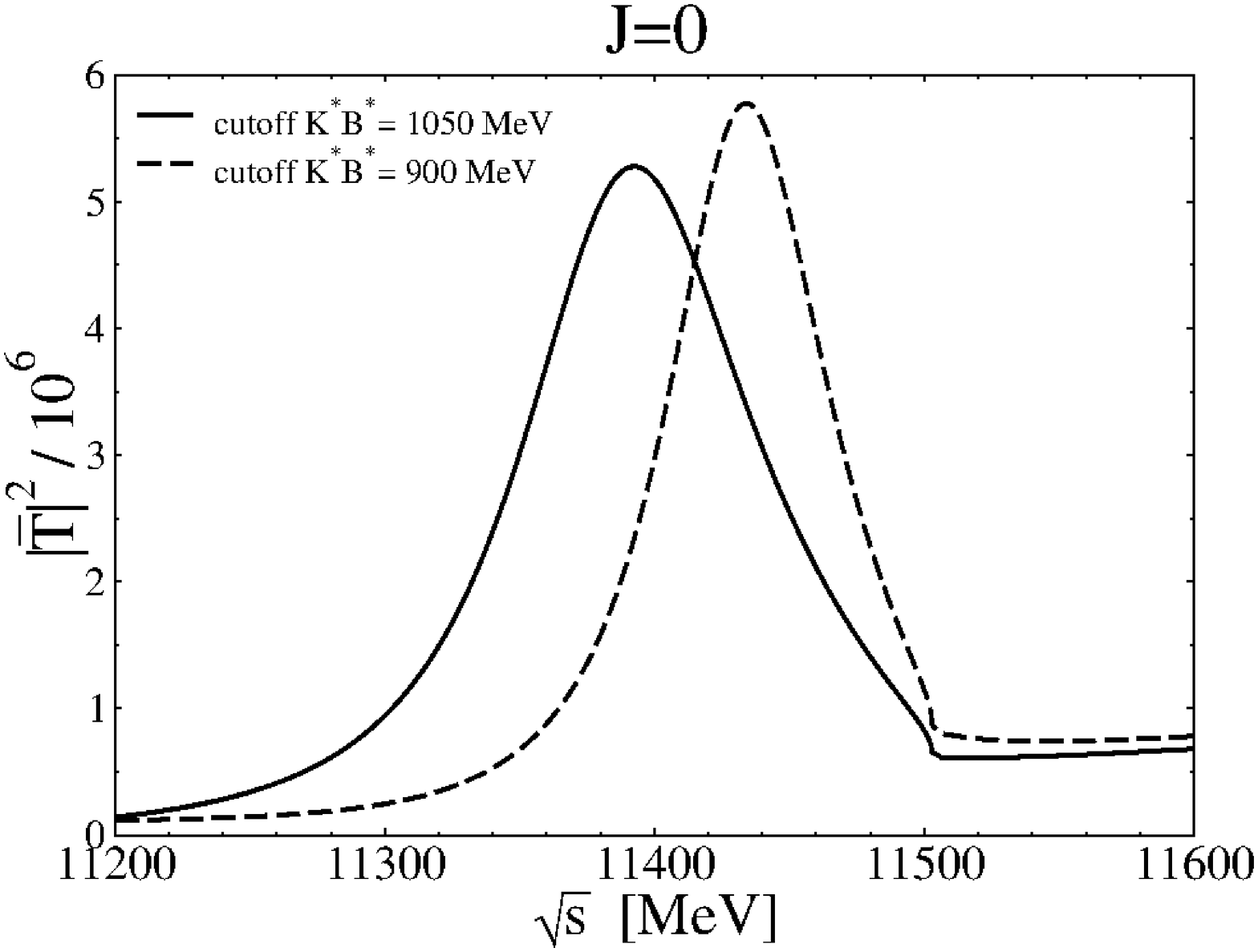}} \\
      \subfigure[]{
          \includegraphics[width=.88\linewidth]{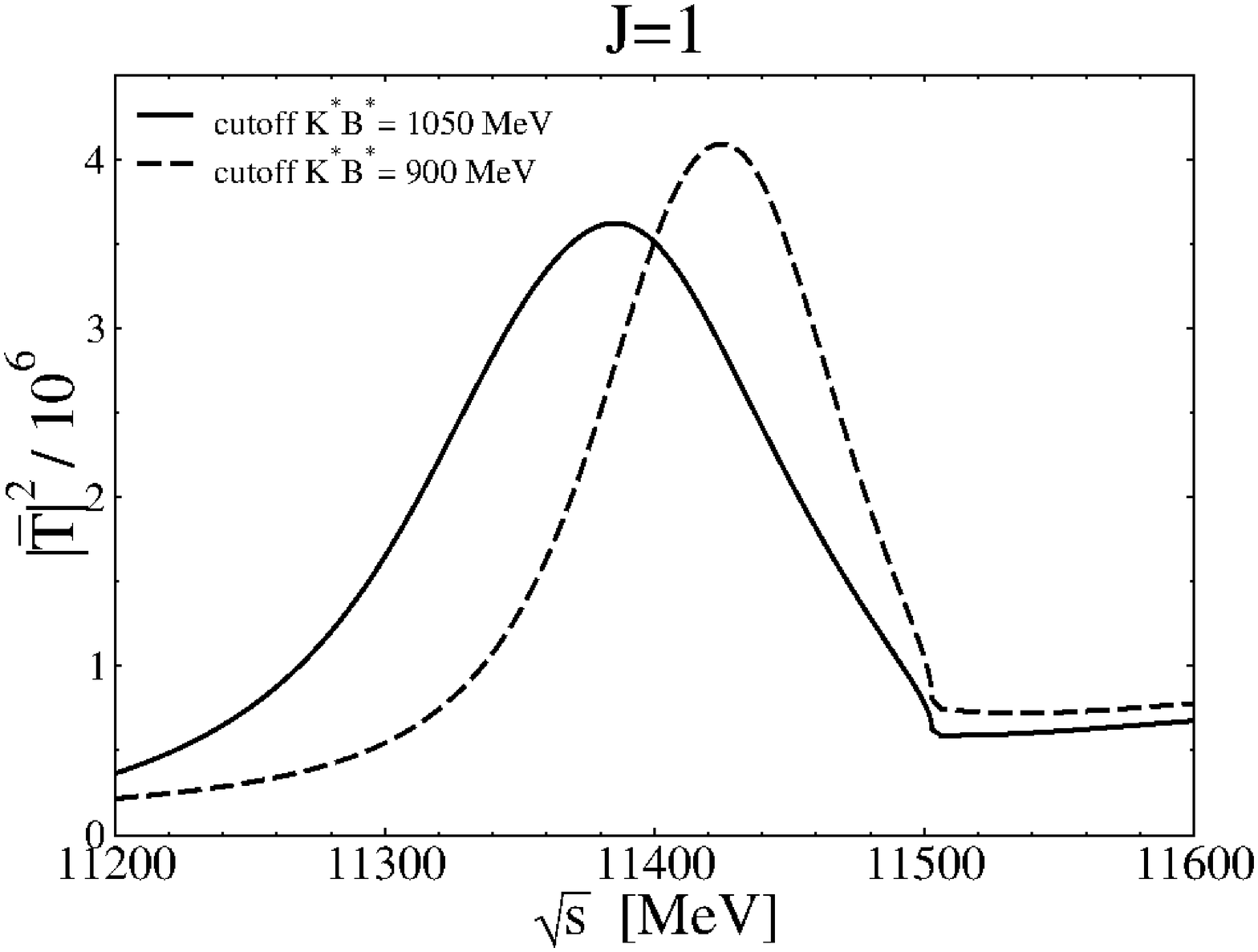}} \\
       \subfigure[]{
          \includegraphics[width=.88\linewidth]{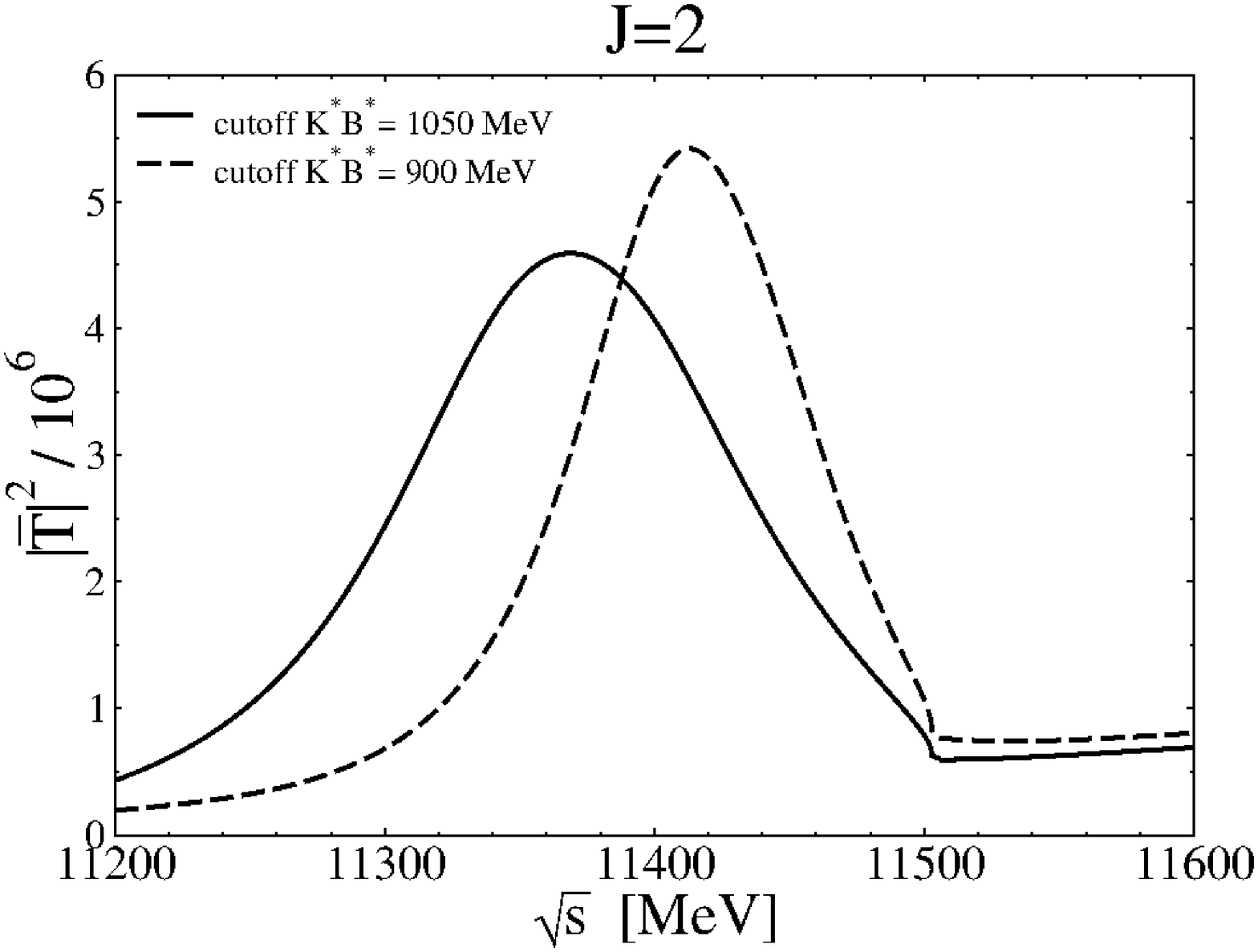}}
   \caption{The three-body amplitude $|\bar T|^2$ as a function of the three-body invariant mass energy, $\sqrt{s}$, for the three different values of the total spin, $J$, and for
 $\qmax^{\barks\barbs}=900$~MeV and $\qmax^{\barks\barbs}=1050$~MeV.}
     \label{fig:res1}
\end{figure}
 
In Fig.~\ref{fig:res1}  we show the results for the three body scattering amplitudes $|\bar T|^2$ as a function of the  total invariant mass energy, $\sqrt{s}$, for the  three possible values of the global spin, $J=0$, 1 and 2. We show the calculations for the extreme values, $900\mev$ and $1050\mev$, of the range considered for $\qmax^{\barks\barbs}$. The difference between these results can be considered as an estimation of the largest uncertainty of our calculation. The consideration of the value of  $\qmax$ for $\barbs\barbs$ in the range $400-450\mev$ has an effect of a shift in the peaks of less than $10\mev$ which is smaller than the uncertainty from $\qmax^{\barks\barbs}$ and hence we show the results only for an intermediate value of $\qmax=420\mev$. Note, however, that the uncertainty from these cutoffs comes inherited from the two-body amplitudes and thus it is not  genuine of the three body model.

For $J=0$ we can see a sharp peak at $\sqrt{s}=11393$~MeV, for  $\qmax^{\barks\barbs}=1050\mev$,
which is
below the $\barks[\barbs\barbs]$ correlated threshold, $11503$~MeV.  This peak can thus be considered as a three-body 
$\barks\barbs\barbs$ bound state, with a binding energy of about $150~\mev$ defined from the uncorrelated threshold  $m_\barks+2m_\barbs= 11543\mev$. Out of this energy,  $40\mev$ comes from the binding of the $\barbs\barbs$ cluster \cite{Dai:2022ulk}, and the rest comes from the three body dynamics.

\begin{table}[t]
 \centering
 \begin{tabular}[t]{c|c|c}
 \hline\hline
  &~$E_B$ &$ \Gamma$ \\ \hline
  $J=0$~&~109--150 & 72--104 \\
  $J=1$~&~118--158&106--153 \\
  $J=2$~&~130--174&103--149 \\
  \hline\hline
 \end{tabular}\\\vspace{3mm}
 \caption{
 Binding energy, $E_B$, and width, $\Gamma$, of the $\barks\barbs\barbs$ system for the three different possible total spin $J$.
The first number in the numerical cells represent the value obtained with the  cutoff $\qmax^{\barks\barbs}=900$~MeV and the second one using $1050\mev$. All units are MeV. The binding energies are referred to the uncorrelated $\barks\barbs\barbs$ threshold,  $m_\barks+2m_\barbs= 11543\mev$
 }
\label{table:results}
\end{table}
 
The results, also for $J=1$ and $J=2$, are summarized in  Table~\ref{table:results}, where we show
the binding energies, $E_B$, and the widths obtained for the two values of $\qmax^{\barks\barbs}$ considered. The binding energies shown in the Table are defined as the difference between the position of the maximum of the peak and the uncorrelated threshold $m_\barks+2m_\barbs= 11543\mev$.

According to Eq.~\eqref{eq:t1IJ} the  single-scattering three-body $J=0$ amplitude
is proportional to the two-body $J=1$ one.
Therefore, the origin of the three-body peak found for total $J=0$ can be traced to the bound $\barks\barbs$ $I(J^P)=0(1^+)$ state which has a mass of $6113\mev$, (see Table I in Ref.~\cite{Oset:2022xji}). This value for the two-body energy, $\sqrt{s_1}$, corresponds, using Eq.~\eqref{eq:s1}, to  a three-body energy $\sqrt{s}=11392\mev$, which is almost where the three-body  state is located. This is an indication that the multiple scattering (Fig.~\ref{fig:FCA}b and c) is small since, if this were the case, we could neglect $G_0$  in Eq.~\eqref{eq:barTfinal} and then $\bar T\simeq 2\bar t_1$.

\begin{figure}[!t]
\begin{center}
\includegraphics[width=0.99\linewidth]{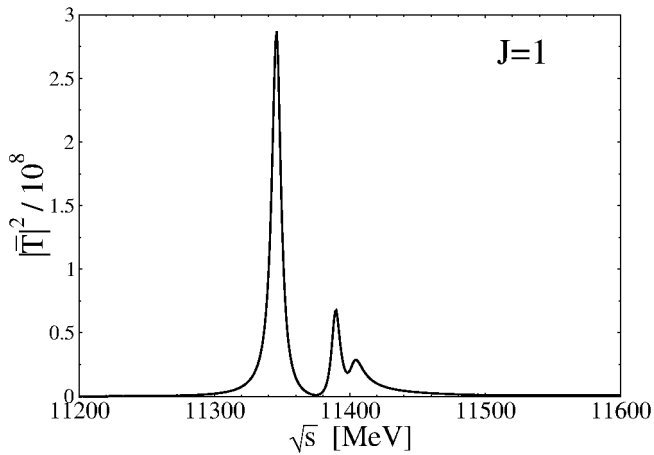}
\caption{Three-body amplitude $|\bar T|^2$ for $J=1$ and 
 $\qmax^{\barks\barbs}=1050$~MeV reducing artificially the main source of imaginary part of the $\barks\barbs$ amplitudes to 5\% its true value.}
\label{fig:T2nowidth}
\end{center}
\end{figure}

The state found for $J=1$ and $J=2$ can be traced to the $\barks\barbs$ two-body states following a similar argument as done above for $J=0$. For $J=1$ the three body amplitude depends on a combination of the three possible two-body spin amplitudes, and for $J=2$ on the three two-body spins, Eq.~\eqref{eq:t1IJ}. According to the results of Ref.~\cite{Oset:2022xji}  (see Table I in Ref.~\cite{Oset:2022xji})
three $\barks\barbs$ states were found with energies $6125\mev$ ($J=0$), $6113\mev$ ($J=1$) and $6074\mev$ ($J=2$) and widths $160\mev$, $98\mev$ and $138\mev$ respectively. These energies correspond to $\sqrt{s}= 11405\mev$, $11392\mev$ and $11351\mev$ respectively, from Eq.~\eqref{eq:s1}. Therefore, up to effects of the non-resonant isospin $I=1$ amplitudes, and the multiple scattering mechanisms, the three-body pole found is essentially the effect of an overlap of these three states due to their large width. Indeed if, for illustrative purposes, we artificially reduced the main source of imaginary part of the $\barks\barbs$ amplitudes, which are the box diagrams with intermediate $\bark \barb$ and $\bark \barbs$ \cite{Oset:2022xji}, to 5\% of its true value, then we would see three clear narrow peaks in the three-body amplitudes, as is shown in Fig.~\ref{fig:T2nowidth} for the $J=1$ case. The $J=2$ case is qualitatively analogous.

Another point worth commenting is that in \cite{Ikeno:2022jbb} the bound states for the $ D^* D^* \bar{K}^*$ follow a similar behaviour like ours, as far as the number of poles found and its nature is concerned, but they are less bound than what we obtain for the $\barks\barbs\barbs$ system.
This trend of the binding energy with the heavy meson mass has also been commonly observed in other studies  when passing from the charm to the bottom sector \cite{Ader:1981db,Ke:2021rxd,Carlson:1987hh,Zouzou:1986qh}.

\section{Summary}

We have studied theoretically the three-body system
 $\barks\barbs\barbs$ to look for possible mesonic states  with
open $s$ and two $b$ flavors.
The work is motivated by the results of a previous work where the  $\barbs\barbs$ with $I(J^P)=0(1^+)$ was found to bind, and, in other work the $\barks\barbs$ interaction was also found to be attractive in $I=0$. This allows us to apply the fixed center approximation (FCA)
to the Faddeev equations where the $\barks$ interacts with each of the $\barbs$ in the $\barbs\barbs$ cluster and undergoes multiple rescattering.
The total three-body amplitude can then be written algebraically in terms of the two-body $\barks\barbs$ obtained from the unitarization of interacting potentials obtained from suitable extensions to the heavy bottom of local hidden gauge symmetry Lagrangians.
The method contains no further degrees of freedom besides the uncertainties already implied in the two body scattering amplitudes.

We find resonant three-body structures with quantum numbers $I(J^P)= 1/2(0^-)$, $1/2(1^-)$ and  $1/2(2^-)$ with binding energies and widths of the order of hundred MeV (see Table~\ref{table:results} and Fig.~\ref{fig:res1}). We hope that these super-exotic mesons, with open strange and double-bottom flavor, can be experimentally found in a not very far future.

\section*{Acknowledgments}
We thank Eulogio Oset for reading a draft version of the manuscript.
The work of N. I. was partly supported by JSPS KAKENHI Grant Numbers JP19K14709 and JP21KK0244.
L.R. acknowledges support by the 
Spanish Ministerio de Ciencia e Innovaci\'on (MICINN)
and the European Regional Development Fund (ERDF) under contract PID2020-112777GB-I00.



\end{document}